\documentclass[10pt]{article}
\usepackage{euscript,amssymb,amsmath}
\usepackage[dvips]{graphicx}
\catcode`\@=11

\newcommand{\be}[1]{\begin{equation}\label{#1}}
\newcommand{\ee}{\end{equation}}
\newcommand{\ba}[1]{\begin{eqnarray}\label{#1}}
\newcommand{\ea}{\end{eqnarray}}
\newcommand{\rf}[1]{(\ref{#1})}
\newcommand{\nn}{\nonumber}
\newcommand{\p}{\partial}
\newcommand{\etal}{{\it et al }}

\newcommand{\const}{\mbox{\rm const}\,}

\def\RR{\mathbb{R}}
\def\NN{\mathbb{N}}

\newcommand{\cD}{\mathcal{D}}
\newcommand{\cH}{\mathcal{H}}

\newcommand{\cK}{\mathcal{K}}
\newcommand{\cP}{\mathcal{P}}
\newcommand{\cT}{\mathcal{T}}

\newcommand{\fu}{\mathfrak{u}}

\newcommand{\fA}{\mathfrak{A}}
\newcommand{\fB}{\mathfrak{B}}

\def\p{\partial}
\def\a{\alpha}

\def\g{\gamma}
\def\e{\epsilon}

\def\lb{\lambda}

\title{The spherically symmetric $\a^2-$dynamo and some of its spectral peculiarities}
\author{Uwe G\"unther$^a$\thanks{e-mail: u.guenther@fzd.de}, Oleg N. Kirillov$^b$\thanks{e-mail: kirillov@imec.msu.ru},
Boris F. Samsonov$^c$\thanks{e-mail: samsonov@phys.tsu.ru} \ and
Frank
Stefani$^a$\thanks{e-mail: f.stefani@fzd.de}\\[2ex]
$a$: Research Center Dresden-Rossendorf,\\  POB 510119, D-01314
Dresden, Germany\\[1ex]
$b$: Moscow State Lomonosov University, Institute of Mechanics,\\
Michurinskii pr. 1, 119192 Moscow, Russia\\[1ex]
$c$: Physics Department, Tomsk State University,\\ 36 Lenin Avenue,
634050 Tomsk, Russia %\\[1ex]
}
\date{13.03.2007}
\begin{document}
\maketitle

\begin{abstract}
A brief overview is given over recent results on the spectral
properties of spherically symmetric MHD $\a^2-$dynamos. In
particular, the spectra of sphere-confined fluid or plasma
configurations with physically realistic boundary conditions (BCs)
(surrounding vacuum) and with idealized BCs (super-conducting
surrounding) are discussed. The subjects comprise third-order branch
points of the spectrum, self-adjointness of the dynamo operator in a
Krein space as well as the resonant unfolding of diabolical points.
It is sketched how certain classes of dynamos with a strongly
localized $\a-$profile embedded in a conducting surrounding can be
mode decoupled by a diagonalization of the dynamo operator matrix. A
mapping of the dynamo eigenvalue problem to that of a quantum
mechanical Hamiltonian with energy dependent potential is used to
obtain qualitative information about the spectral behavior. Links to
supersymmetric Quantum Mechanics and to the Dirac equation are
indicated.
\end{abstract}

\noindent {\bf Preliminaries}\\ The magnetic fields of planets,
stars and galaxies are maintained by dynamo effects in conducting
fluids or plasmas \cite{mhd-book-1,krause-1,mhd-book-3}. These
dynamo effects are caused by a topologically nontrivial interplay of
fluid (plasma) motions and a balanced self-amplification of the
magnetic fields
--- and can be described within the framework of
magnetohydrodynamics (MHD) \cite{mhd-book-1,krause-1}.

For physically realistic dynamos the coupled system of Maxwell and
Navier-Stokes equations has, in general, to be solved numerically.
For a qualitative understanding of the occurring effects
semi-analytically solvable toy models play an important role. One of
the simplest dynamo models is the so called $\a^2-$dynamo with
spherically symmetric $\a-$profile\footnote{The $\a-$profile $\a(r)$
plays the role of an effective potential for the $\a^2-$dynamo.}
$\a(r)$ (see, e.g. \cite{krause-1}). For such a dynamo the magnetic
fields can be decomposed in poloidal and toroidal components,
expanded over spherical harmonics \cite{krause-1,GS-jmp1} and
unitarily re-scaled \cite{GSZ-squire}. As result one arrives at a
set of $l-$mode decoupled $2\times 2-$matrix differential eigenvalue
problems \cite{krause-1,GS-jmp1,GSZ-squire}
\be{d1}
\fA_\a=\left(\begin{array}{cc}-Q[1] & \alpha\\ Q[\alpha] & -Q[1]
 \end{array}\right),\qquad Q[\alpha]:=-\p_r\alpha(r) \p_r + \alpha(r) \frac{l(l+1)}{r^2}
\ee
with  boundary conditions (BCs) which have to be imposed in
dependence of the concrete physical setup and which will be
discussed below. The $\a-$profile describes the net effect of small
scale helical turbulence on the magnetic field \cite{krause-1}. It
can be assumed real-valued $\a(r)\in\RR$ and sufficiently smooth. We
note that the reality of the differential expression \rf{d1},
independently from the concrete BCs, implies an operator spectrum
which is symmetric with regard to the real axis, i.e. which consists
of purely real eigenvalues and of complex conjugate eigenvalue
pairs.

In \cite{GS-jmp1} it was shown that the differential expression
\rf{d1} of this operator has the fundamental (canonical) symmetry
\cite{azizov,L2}
\be{d2-1} \fA_\a=J\fA_\a^\dagger  J , \qquad
J=\left(\begin{array}{cc}0 & I\\ I & 0
 \end{array}\right)\, .
\ee
In case of BCs compatible with this fundamental symmetry the
operator turns out self-adjoint in a Krein space\footnote{For
comprehensive discussions of operators in Krein spaces see, e.g.,
\cite{azizov,L2,bognar}.}  $(\cK_J,[.,.]_J)$
\cite{GS-jmp1,GSZ-squire} and in this way it behaves similar like
Hamiltonians of $\cP\cT-$symmetric Quantum Mechanics
\nobreak{(PTSQM)} \cite{BB,BBjmp,PT-Z1,most1,BBJ-1,CMB-rev}.
\begin{figure}[htb]                     %instead of \begin{figure}[t]
\begin{center}                         %instead of \begin{center}
\includegraphics[width=0.5\textwidth]{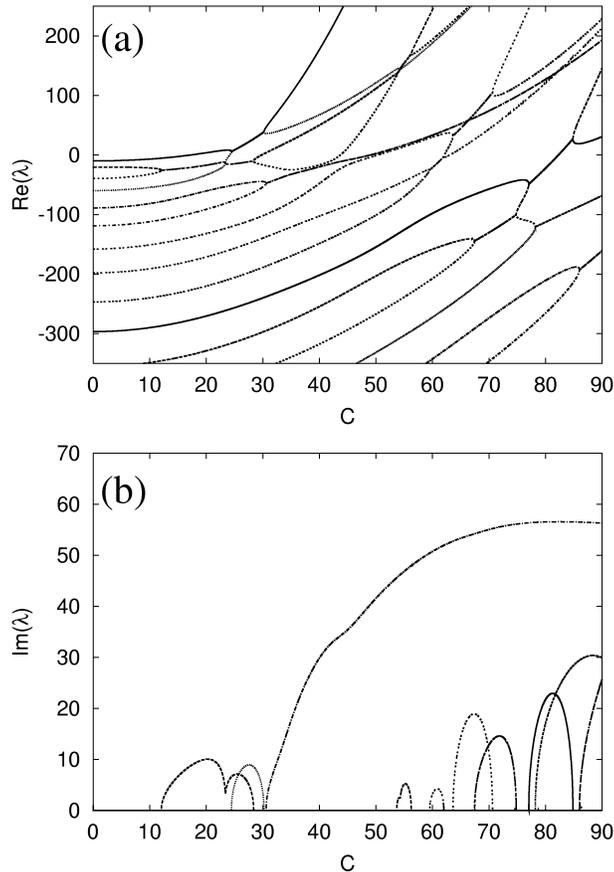}
\end{center}                         %instead of \end{center}
\vspace{-2mm} \caption{Real and imaginary components of the
$\alpha^2-$dynamo spectrum as functions of the scale factor $C$ of
an $\alpha-$profile $\alpha(r)=C\times (1-26.09 \times r^2+ 53.64
\times r^3 -28.22 \times r^4)$ in the case of angular mode number
$l=1$ and physically realistic boundary conditions \rf{bc1}. The
concrete coefficients in the quartic polynomial $\alpha (r)$ have
their origin in numerical simulations of  field reversal dynamics
(see Ref. \cite{SG-revers-PRL,EPSL-1}). Only the imaginary
components with $\Im \lambda \ge 0$ are shown. The symmetrically
located complex conjugate $(\Im \lambda \le 0)-$components are
omitted for sake of brevity. \label{fig1}}
\end{figure}

Subsequently, we first present a sketchy overview of some recent
results on the spectral behavior of $\a^2-$dynamos obtained in
\cite{GSZ-squire,GSG-cz2,GS-cz3,GK-jpa2006,KG-PAMM2006,GSS} which we
extend by a discussion of the transition from $\a^2-$dynamo
configurations confined in a box to dynamos living in an unconfined
conducting surrounding.

\noindent {\bf Physically realistic BCs and spectral triple
points}\\
For roughly spherically symmetric dynamical systems like the Earth
the conducting fluid is necessarily confined within the core of the
Earth  so that the $\a-$effect resulting from the fluid motion has
to be confined to this core. Setting the surface of the outer core
at a radius $r=1$  one can assume $\a(r>1)=0$ and a behavior of the
magnetic field at $r>1$ like in vacuum. A multi-pole-like decay of
the magnetic field at $r\to\infty$ leads then to mixed effective BCs
at $r=1$ (see, e.g. \cite{krause-1}) and an corresponding operator
domain of the type
\ba{bc1}
&&\cD(\fA_{\alpha})=\left\{\fu\in \tilde{\cH}=L_2(0,1)\oplus
L_2(0,1)| \ \fu(r\searrow 0)= 0,\ \ \fB\fu|_{r\nearrow
1}= 0\right\},\nn\\
&&\fu:=\left(
        \begin{array}{c}
          u_1 \\
          u_2 \\
        \end{array}
      \right),\qquad \fB:=\left(
                            \begin{array}{cc}
                              \partial_r+\frac lr & 0 \\
                              0 & 1 \\
                            \end{array}
                          \right).
\ea
From the domain $\cD(\fA_{\alpha}^\dagger)$ of the adjoint operator
\ba{bc2}
&&\cD(\fA_{\alpha}^\dagger)=\left\{\hat\fu\in
\hat{\cH}=L_2(0,1)\oplus L_2(0,1)| \ \hat\fu(r\searrow 0)= 0,\ \
\hat\fB\hat\fu|_{r\nearrow 1}= 0\right\},\nn\\
&&\hat\fu:=\left(
        \begin{array}{c}
          \hat u_1 \\
          \hat u_2 \\
        \end{array}
      \right),\qquad \hat\fB:=\left(
                            \begin{array}{cc}
                              \partial_r+\frac lr & -\alpha(r)\partial_r \\
                              0 & 1 \\
                            \end{array}
                          \right)
\ea
one reads off that $\cD(\fA_{\alpha}^\dagger)\neq
\cD(\fA_{\alpha})$ and, hence, the dynamo operator $\fA_\a$ itself
is not self-adjoint even in a Krein space.
\begin{figure}[htb]                    %instead of \begin{figure}[t]
\begin{center}
\includegraphics[width=0.9\textwidth]{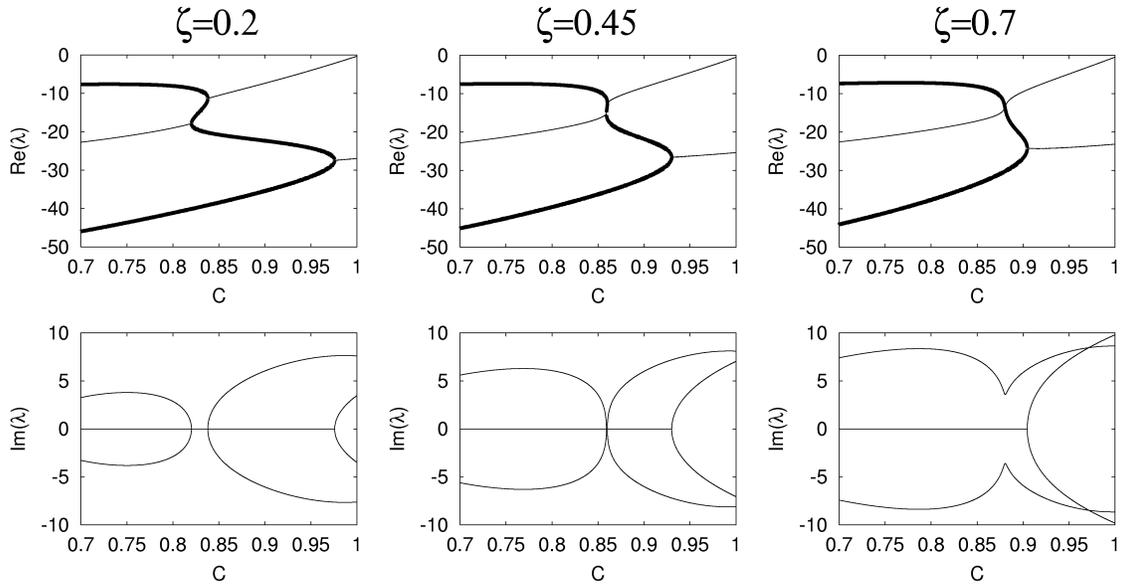}
\end{center}                         %instead of \end{center}
\vspace{-2mm} \caption{ $\alpha^2-$dynamo with $ \alpha(r)=C \left[-
(21.465+2.467 \zeta) + (426.412+ 167.928 \zeta ) r^2-(806.729+
436.289 \zeta ) r^3 \right.$\hfill\mbox{} $\left.+ (392.276+272.991
\zeta ) r^4 \right]$ and a spectral triple point at
$(\zeta=0.45,C=0.86)$. Highlighted (fat) lines correspond to purely
real branches of the spectrum. The cusp in the imaginary component
(lower right graphics) indicates the closely located triple
point.\label{fig2}}
\end{figure}

In case of constant $\a-$profiles and arbitrary $l\in \NN$, the
spectrum is implicitly given by a characteristic equation built from
spherical Bessel functions \cite{krause-1}. In all other cases
numerical studies are required. A typical spectral branch graph is
depicted in Fig. \ref{fig1}. Obviously, for the specific
$\a-$profile it contains a large number of spectral phase
transitions from real spectral branches to complex ones and back.
There are strong indications that phase transition points (second
order branch points/exceptional points) of the spectrum close to the
$\lb=0$ line play an important role in polarity reversals of the
magnetic field (see \cite{SG-revers-PRL,EPSL-1,MHD-pamir,GAFD-2} for
numerical studies and \cite{cadar-EPL} for recent experiments).

Apart from the second-order branch points visible in Fig. \ref{fig1}
there may occur third- and higher-order branch points. They are
located on hyper-surfaces of higher co-dimension in parameter space
and they therefore require a tuning of more parameters to pin them
down\footnote{An explicit hyper-surface parametrization of
second-order branch point configurations embedded in a
$\cP\cT-$symmetric $3\times 3-$matrix model with corresponding
$2\times 2-$Jordan-block preserving modes can be found e.g. in the
recent work \cite{MZ-PLB2007}.}. Corresponding results have been
obtained in \cite{GS-cz3} and are illustrated in Fig. \ref{fig2}.
The triple points result from coalescing second-order branch points,
correspond to $3\times 3$ Jordan blocks in the spectral
decomposition of the operator and are accompanied by a merging or
disconnecting of two complex spectral sectors over the parameter
space. An implicit indication of a closely located triple point is
the presence of cusps in the imaginary components as they are
visible in Figs. \ref{fig1}, \ref{fig2}.

\noindent {\bf Idealized BCs and Krein-space related perturbation
theory}\\
In order to gain some deeper insight into possible dynamo-related
processes semi-analytical toy model considerations play a crucial
role. A certain simplification of the eigenvalue problem has been
achieved in \cite{GK-jpa2006,KG-PAMM2006} by considering a reduced
and idealized (auxiliary) problem\footnote{From a physical point of
view such $\a^2-$dynamos can be regarded as embedded in a
superconducting surrounding.} with Dirichlet BCs imposed at $r=1$,
i.e. by setting $\fu(r=1)=0$.
\begin{figure}[htb]
\begin{center}
  \includegraphics[width=\textwidth]{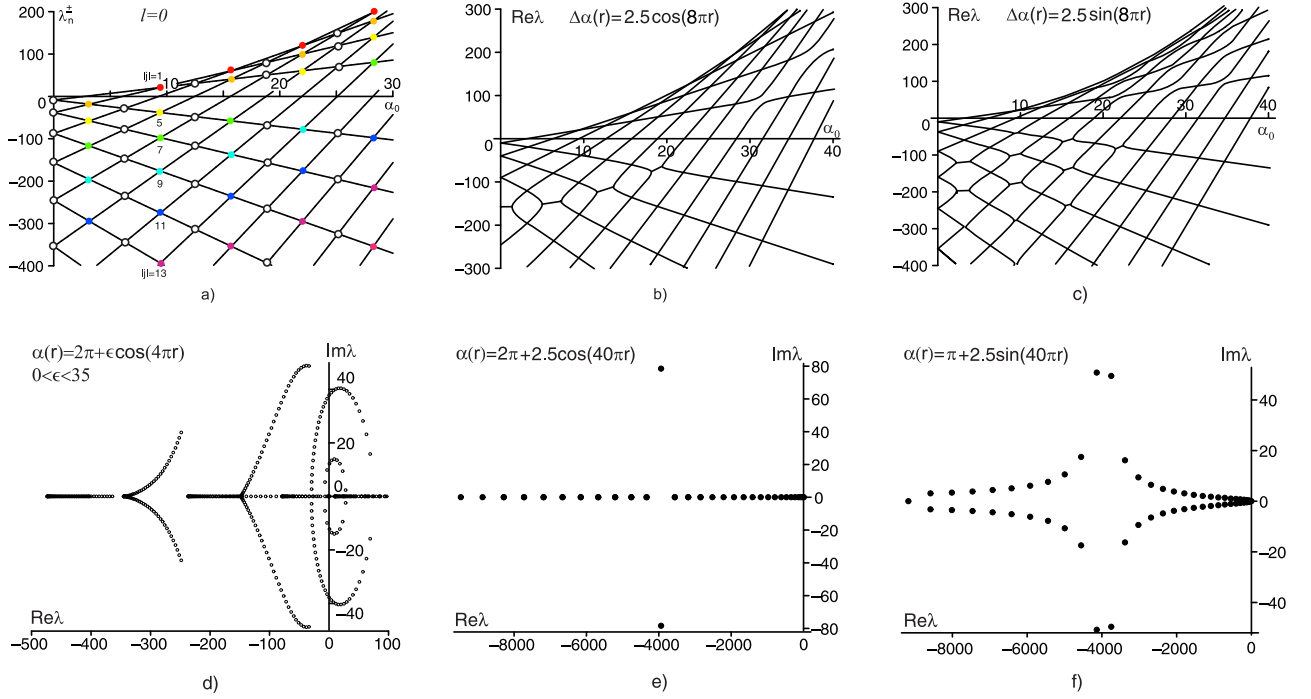}
\end{center}                         %instead of \end{center}
\vspace{-2mm} \caption{The spectral mesh of the operator matrix
$\fA_{\alpha}$ for $l=0$ (a); its resonant deformation due to
harmonic perturbations of a constant $\alpha$-profile (b), (c); the
formation of overcritical oscillatory dynamo regimes for $\epsilon$
increasing from $\epsilon=0$ ($\Im \lambda=0$) to $\epsilon=35$
($\Re \lambda
>0$, $\Im \lambda\neq 0$ for some branches) (d); and the resonant
unfolding of DPs in the complex plane (e), (f).} \label{fig3}
\end{figure}
In this case it holds $\cD(\fA_\a)=\cD(\fA_\a^\dagger)$ and the
operator $\fA_\a$ is self-adjoint in a Krein space $(\cK_J,[.,.]_J)$
\cite{GSZ-squire}. For constant $\a-$profiles
$\alpha(r)=\alpha_0=\const$ the eigenvalue problem
$(\fA_{\a_0}-\lb)\fu=0$ becomes exactly solvable in terms of
orthonomalized Riccati-Bessel functions
\be{bc3}
u_n(r)=N_n r^{1/2}J_{l+\frac 12}(\sqrt{\rho_n}r),\qquad
N_n:=\frac{\sqrt 2 }{ J_{l+\frac 32}(\sqrt{\rho_n})},\qquad
(u_m,u_n)=\delta_{mn},\qquad \| u_n\|=1
\ee
with $\rho_n>0$ the squares of Bessel function roots
$J_{l+\frac{1}{2}}(\sqrt{\rho_n})=0$. The solutions of the
eigenvalue problem have the form
\be{bc4}
\fu_n^\pm=\left(
\begin{array}{c}
  1 \\
  \pm \sqrt{\rho_n} \\
\end{array}
\right)u_n\in \RR^2\otimes L_2(0,1)\,,
\ee
are Krein space orthonormalized
\be{bc5}
[\fu_m^\pm,\fu_n^\pm ]=\pm 2 \sqrt{\rho_n}\delta_{mn},\qquad
[\fu_m^\pm,\fu_n^\mp]=0, \qquad \fu_n^\pm\in \cK_\pm\subset
\cK,\qquad \fu_n^\pm=:\fu_n^\varepsilon, \ \varepsilon=\pm
\ee
and correspond to eigenvalue branches
$\lambda_n^{\varepsilon}=-\rho_n+\varepsilon \alpha_0 \sqrt{\rho_n}
$ which  scale linearly with $\a_0$. In the $(\a_0,\Re\lb)-$plane
the branches $\lb_n^+$ and $\lb_n^-$ of states $\fu^+_n$, $\fu_n^-$
of positive and negative Krein space type form a spectral mesh (see
Fig. \ref{fig3}).  The intersection points (nodes of the mesh) are
semisimple double eigenvalues, i.e. eigenvalues of geometrical and
algebraical multiplicity two --- so called diabolical points (DPs)
\cite{berry-3}. Two given branches $\lb_n^\varepsilon(\a_0)$ and
$\lb_m^\delta(\a_0)$ intersect at the single point
$\lambda=\lb^\nu_0:=\varepsilon\delta\sqrt{\rho_n\rho_m}$, \
$\a_0=\a_0^\nu:=\varepsilon\sqrt{\rho_n}+\delta\sqrt{\rho_m}$ and
one obtains that branches from states of opposite Krein space type
$\varepsilon=-\delta$ intersect for $\lb_0^\nu<0$, whereas states of
the same type $(\varepsilon=\delta)$ intersect at $\lb_0^\nu>0$.
Under small inhomogeneous perturbations
$\a(r)=\a_0^\nu+\Delta\alpha(r)=\a_0^\nu+\e\phi(r)$ the diabolical
points split $\lambda_0^\nu\mapsto \lb_0^\nu+\e\lb_1+\ldots$ into
two real or complex points (see also \cite{KS} for similar
considerations) with leading contribution $\lb_1$ resulting from the
quadratic equation
\be{bc6}
\lambda_1^2- \lambda_1\left(\varepsilon\frac{[\fB
\fu_n^{\varepsilon}, \fu_n^{\varepsilon} ]}{2\sqrt{\rho_n} }+
\delta\frac{[\fB \fu_m^{\delta}, \fu_m^{\delta} ]}{2\sqrt{\rho_m}
}\right)+\varepsilon\delta\frac{[\fB \fu_n^{\varepsilon},
\fu_n^{\varepsilon} ][\fB \fu_m^{\delta}, \fu_m^{\delta} ] -[\fB
\fu_n^{\varepsilon}, \fu_m^{\delta}
]^2}{4\sqrt{\rho_n\rho_m}}=0\,,
\ee
where
\be{bc7}
[\fB \fu_m^{\delta}, \fu_n^{\varepsilon}]
=\int_0^1\varphi\left[\left(\varepsilon\delta\sqrt{\rho_n\rho_m}+\frac{l(l+1)}{r^2}\right)u_mu_n+u_m'u_n'\right]dr.
\ee
The unfolding of the DPs follows the typical Krein space rule. When
they result from branches of the same type $(\varepsilon=\delta,
\lb_0^\nu>0)$ then the corresponding DPs unfold purely real-valued,
whereas DPs from branches of  opposite type $(\varepsilon=-\delta,
\lb_0^\nu<0)$ may unfold into complex conjugate eigenvalue pairs.
This behavior is clearly visible in Fig. \ref{fig3} b,c. Direct
inspection reveals  that the spectral meshes of unperturbed
operators $\fA_{\a_0}$ for $l=0$ and $0< l\ll\infty$ show strong
qualitative similarities so that results obtained for the
quasi-exactly solvable $(l=0)-$model will qualitatively hold for
models with $0< l\ll\infty$ too. Via Fourier expansion of $\a(r)$ a
very pronounced resonance has been found along parabolas in the
$(\a_0,\Re\lb)-$plane indicated by white and colored dots in Fig.
\ref{fig3}a --- leaving regions away from these parabolas almost
unaffected. An especially pronounced resonance is induced by cosine
perturbations which in linear approximation affect only the single
parabola $j=2k$, Fig. \ref{fig3}b,e. Sine perturbations act
strongest on parabolas $|j|=2k\pm 1$ with decreasing effect on
$|j|=2k\pm m$ for increasing $m$ (see Fig. \ref{fig3}c,f).
Physically important is the fact that higher mode numbers $k$
(shorter wave lengths of the $\Delta\a(r)$ perturbations) affect
more negative $\Re\lb$. Due to a magnetic field behavior $\propto
e^{\lb t}$ this is the mathematical formulation of the physically
plausible fact that small-scale perturbations decay faster than
large-scale perturbations. Numerical indications for the importance
of this behavior in the subtle interplay of polarity reversals and
so called excursions ("aborted" reversals) of the magnetic field
have been recently given in \cite{GAFD-2}.

 \noindent {\bf Diagonalizable $\a^2-$dynamo operators,
SUSYQM and
the Dirac equation}\\
Another approach to obtain quasi-exact solution classes of the
eigenvalue problem $(\fA_\a-\lb)\fu=0$ consists in a
$\lb-$dependent diagonalization of the operator matrix \rf{d1}.
The basic feature of this technique, as demonstrated in
\cite{GSS}, is a two-step procedure consisting of a gauge
transformation which diagonalizes the kinetic term and a
subsequent global (coordinate-independent) diagonalization of the
potential term. Such an operator diagonalization is possible for
$\a-$profiles satisfying the constraint
\be{bc8}
\a''(r)+\frac 12\a^3(r)-a^2\a(r)=0
\ee
with $a=\const\in\RR$ a free parameter. Solutions $\a(r)$ of this
autonomous differential equation (DE) can be expressed in terms of
elliptic integrals. In order to maximally explore similarities to
known QM type models\footnote{For early comments on structural links
between MHD dynamo models and QM-related eigenvalue problems see
e.g. \cite{meinel}.} a strongly localized $\a-$profile has been
assumed which smoothly vanishes toward $r\to \infty$. Physically,
such a setup can be imagined as a strongly localized
dynamo-maintaining turbulent fluid/plasma motion embedded in an
unbounded conducting surrounding (plasma) with fixed homogeneous
conductivity. The only $\a-$profile with $\a(r\to\infty)\to 0$
satisfying \rf{bc8} has the form of a Korteweg-de Vries(KdV)-type
one-soliton potential
\be{bc9}
\alpha(r)=\frac{2a}{\cosh [a(r-r_0)]}\,.
\ee
This amazing finding indicates on deep structural links to KdV and
supersymmetric quantum mechanics (SUSYQM) and opens up a completely
new exploration approach to $\a^2-$dynamos\footnote{The question of
whether this new class of quasi-exactly solvable $\a^2-$dynamo
models might be structurally related (via dynamical embedding) to
the recently studied $\cP\cT-$symmetrically extended KdV solitons
\cite{CMB-PTsolit-1,fring-PTsolit} remains to be clarified.}. In
\cite{GSS} we restricted the consideration to the most elementary
solution properties of such models. The decoupled equation set after
a parameter and coordinate rescaling has been found in terms of two
quadratic pencils
\be{bc10}
[-\p_x^2+\frac{l(l+1)}{x^2}-\frac 12\a^2+\frac
12\mp\e\a-\e^2]F_\pm=0\,, \quad \a=\frac{2}{\cosh(x-x_0)}
\ee
in the new variable $x:=ar$ and with new auxiliary spectral
parameter $\e=\left(\frac12 -\lambda\right)^{1/2}$. The equivalence
transformation from $(\fA_\a-\lb)\fu=0$ to \rf{bc10} is regular for
$\e\neq 0$ and becomes singular at $\e=0$ where \rf{bc10} has to be
replaced by a Jordan type equation system
\be{bc11}
\left(
  \begin{array}{cc}
    \partial_x^2-V_0 & -V_1 \\
    0 & \partial_x^2-V_0 \\
  \end{array}
\right)\left(
         \begin{array}{c}
           \Xi_1 \\
           \Xi_0 \\
         \end{array}
       \right)=0
\ee
with potentials  $V_0=l(l+1)x^{-2}-\frac 12(\a^2-1)$, $V_1=-\a$. In
terms of the original spectral parameter $\lb$ the eigenvalue
problems \rf{bc10} read
\be{bc10-2}
\left[-\partial_x^2 +\frac{l(l+1)}{x^2} -\frac 12 \alpha^2\mp
\left(\frac 12-\lb\right)^{1/2}\alpha\right]F_\pm =-\lambda F_\pm
\ee
and can be related to the spectral problem of a QM Hamiltonian with
energy $E=-\lb$ and energy-dependent potential component $\mp
\left(\frac 12-\lb\right)^{1/2}\a(x)=\mp \left(E+\frac
12\right)^{1/2}\a(x)$. For physical reasons asymptotically vanishing
field configurations with $F_\pm (x\to \infty)\to 0$,
$\Xi_{0,1}(x\to \infty)\to 0$ are of interest. These Dirichlet BCs
at infinity imply the self-adjointness of the operator $\fA_\a$ in a
Krein space $\cK_J$
--- with \rf{bc10}, \rf{bc11} as  special representation of the
eigenvalue problem $(\fA_\a-\lb)\fu=0$. From the structure of
\rf{bc10},\rf{bc10-2} follows that the only free parameter apart
from the angular mode number $l\in\NN$ is the maximum position $x_0$
of the $\a-$profile $\a(x)$ (the minimum position of the potential
component $-\a^2(x)/2$) so that solution branches will be functions
$\lb(x_0)$.
\begin{figure}[htb]
\begin{center}
\begin{minipage}{0.45\textwidth}
\includegraphics[angle=0, width=0.9\textwidth]{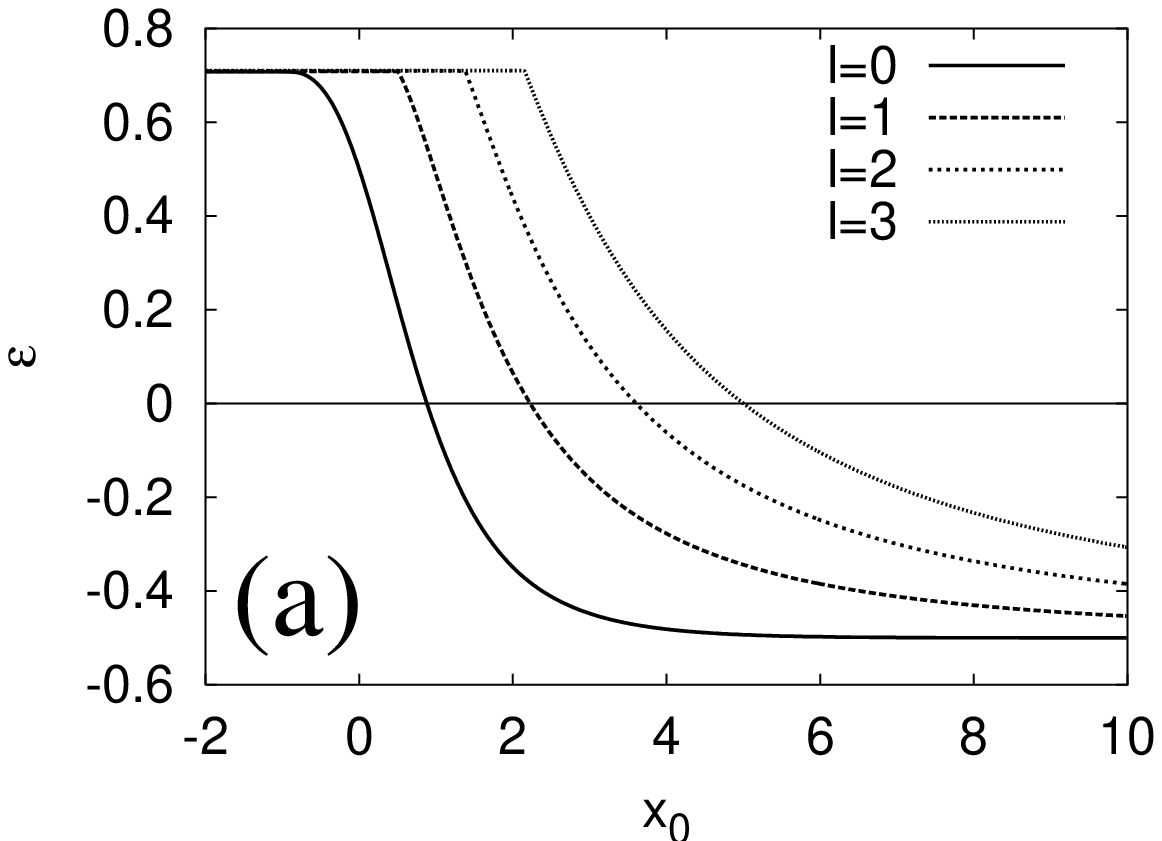}
\end{minipage}
%\hspace{1em} \addtocounter{figure}{-1}
%\renewcommand{\thefigure}%
%{\arabic{figure}b} \hspace{1em}
%\begin{minipage}{5.7cm}
\begin{minipage}{0.45\textwidth}
\includegraphics[angle=0, width=0.9\textwidth]{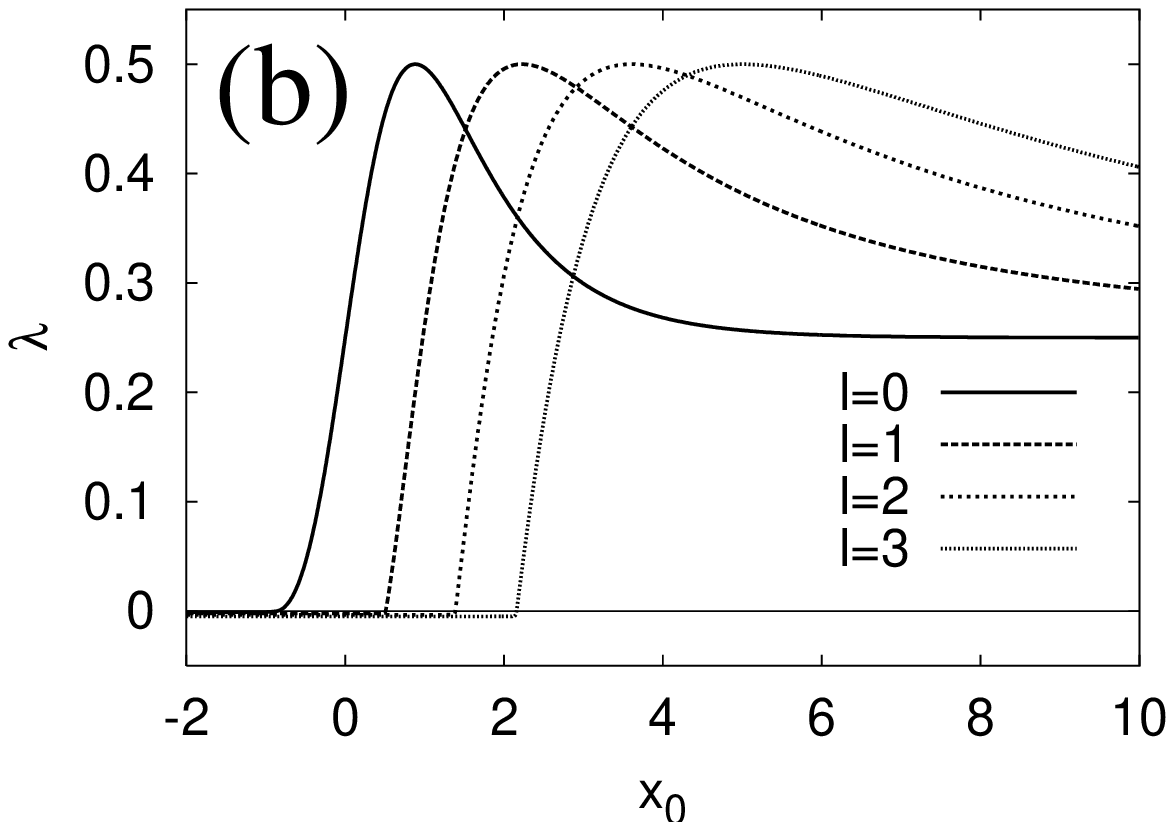}
\end{minipage}
\end{center}
\caption{\label{fig4}\small Spectra $\e(x_0)$ (a) and $\lambda(x_0)$
(b) in case of angular mode numbers $l=0,1,2,3$. For numerical
reasons the Dirichlet BC has been imposed at the large distance
$x=100$.}
\end{figure}

With the help of SUSY techniques it has been shown in \cite{GSS}
that \rf{bc10-2} has a single bound state (BS) type solution which
via $E=-\lb<0$ corresponds to an overcritical dynamo mode $\lb>0$.
It has been found that the BS solutions of \rf{bc10-2} behave
differently for $x_0<x_J$  and $x_0>x_J$, where for $x_0=x_J$ the
description in terms of \rf{bc10} breaks down and has to be replaced
by the singular Jordan type representation \rf{bc11}. By a SUSY
inspired factorization ansatz
\ba{sol11}
&&-\p_x^2+\frac{l(l+1)}{x^2}-\frac12\alpha^2+\frac12=L^\dagger L\,,\label{sol11-a}\\
&&L=-\p_x+w,\quad L^\dagger=\p_x+w,\quad w=u'/u\,,\label{sol11-b}
\ea
an equivalence relation between \rf{bc10} and a system of two
Dirac equations
\be{DS}
H_\pm\Psi_\pm=\e\Psi_\pm\,,\quad H_\pm=\g\p_x+V_\pm
\ee
\be{DH}
\Psi_\pm=\left(
\begin{array}{c}
\psi_{1,\pm}\\\psi_{2,\pm}
\end{array}
\right):=\left(
\begin{array}{c}
F_\pm\\\e^{-1}LF_\pm
\end{array}
\right),\quad \g=\left(
\begin{array}{cc}
0 & 1\\-1 & 0
\end{array}
\right),\quad V_\pm=\left(
\begin{array}{cc}
\mp\a & w\\w & 0
\end{array}
\right)
\ee
has been established for models with $x_0<x_J$. General results on
Dirac equations allowed then for the conclusion that in case of
$x_0<x_J$ the bound state related spectrum has to be real. A
perturbation theory with the distance $\delta=x_0-x_J$ from the
Jordan configuration as small parameter supplemented by a bootstrap
analysis showed that the Dirichlet BCs $F_\pm(x\to\infty)\to 0$
render only the solution $F_+(x)$ non-trivial and with real
eigenvalue, whereas $F_-(x)$ has to vanish identically $F_-(x)\equiv
0$. The single spectral branch in terms of $\lb(x_0)$ and $\e(x_0)$
is depicted in Fig. \ref{fig4} for angular mode numbers $l=0,1,2,3$.

Assuming the dynamo model with strongly localized $\a-$profile
\rf{bc9}, \rf{bc10} confined in a large box, i.e. with Dirichlet BCs
imposed at large $x=X\gg0$, one can study the  dynamo spectrum in
the infinite box limit.
\begin{figure}[htb]
\begin{center}
\includegraphics[angle=0, width=0.9\textwidth]{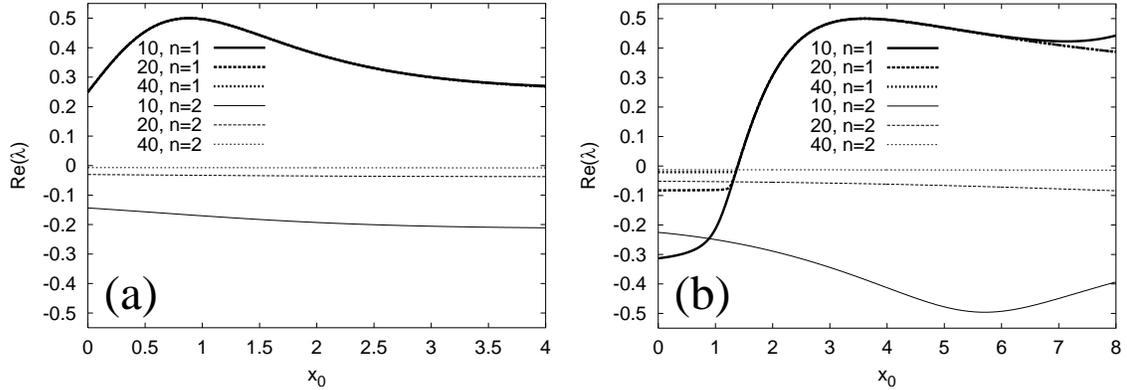}
\end{center}
\caption{\label{fig5}\small Cutoff ($X$-)dependence of the spectral
branches with radial mode numbers $n=1,2$ and angular mode numbers
$l=0$ (a) and $l=2$ (b) for cutoffs (box-lengths) $X=10,\, 20,\,
40$. Clearly visible are the $X-$independence of the overcritical BS
type modes $(n=1)$ and the tendency $\lb\propto -1/X^2$ for the
undercritical (box-type) mode. The modes with $n\ge 3$ show the same
qualitative $\lb\propto -1/X^2$ behavior like the $n=2$ mode and are
note depicted here.}
\end{figure}
Figures \ref{fig5} a,b show the corresponding behavior. Due to its
localization the BS-related overcritical dynamo mode  is almost
insensitive  to the $X\to\infty$ limit. This is in contrast to the
under-critical (decaying) modes which behave as expected for a sign
inverted box spectrum of QM. For fixed mode number $n\ge 2$ and
$X\to\infty$ the energies $E_n$ decrease  like $E_n\propto
1/X^2\searrow 0$ and the corresponding part of the spectrum becomes
quasi-continuous and related to the continuous (essential) spectrum
of QM scattering states of a particle moving in the energy dependent
potential $\frac{l(l+1)}{x^2} -\frac 12 \alpha^2(x)\mp \left(E+\frac
12\right)^{1/2}\alpha(x)$. For the associated dynamo eigenvalues
this implies $\lb_n\propto -1/X^2\nearrow 0$ --- as it is clearly
visible in Figures \ref{fig5} a,b.

\noindent {\bf Concluding remarks}\\
A brief overview over some recent results on the spectra of dynamo
operators has been given. The obtained structural features like the
resonance effects in the unfolding of diabolical points as well as
the unexpected link to KdV soliton potentials, elliptic integrals,
SUSYQM and the Dirac equation appear capable to open new
semi-analytical approaches to the study of $\a^2-$dynamos.

\noindent {\bf Acknowledgement}\\
The work reviewed here has been supported by the German Research
Foundation DFG, grant GE 682/12-3 (U.G.), by the CRDF-BRHE program
and the Alexander von Humboldt Foundation (O.N.K.) as well as  by
RFBR-06-02-16719 and SS-5103.2006.2 (B.F.S.).

\end{document}